\documentclass[10pt]{article}

\usepackage{amsmath}
\usepackage{amssymb}
\usepackage[section]{placeins}
\usepackage{graphicx}
\usepackage{cite}
\usepackage{color}

\usepackage[labelfont=bf,labelsep=period,justification=raggedright]{caption}

\makeatletter
\renewcommand{\@biblabel}[1]{\quad#1.}
\makeatother

\date{}

\newcommand{\dnn}{d}
\newcommand{\dnnmedian}{\widetilde{\dnn}}
\newcommand{\mm}{MM}
\newcommand{\ms}{MS}
\newcommand{\sm}{SM}
\newcommand{\sstat}{SS}
\newcommand{\prob}[1]{P_{#1}}
\newcommand{\step}{\ell}
\newcommand{\turn}{\theta}
\newcommand{\dir}{\Theta}
\newcommand{\turnspread}{\rho}

\newcommand{\ann}{\phi}

\newcommand{\matlab}{M{\sc atlab}}
\newcommand{\inter}{{INT}}
\newcommand{\nonint}{{NON}}
\newcommand{\mediandiff}{\Delta \widetilde{x}}
\renewcommand{\exp}[1]{\mathrm{e}^{#1}}
\newcommand{\fracmoving}{M_\%}
\newcommand{\experiment}{EXP}

\begin{document}

\begin{flushleft}
{\Large
\textbf{Social aggregation in pea aphids: Experimental measurement and stochastic modeling}
}
\\
Christa Nilsen$^{1}$, 
John Paige$^{1}$, 
Olivia Warner$^{1}$, 
Benjamin Mayhew$^{1}$, 
Ryan Sutley$^{1}$, 
Matthew Lam$^{2}$, 
Andrew J. Bernoff$^{2}$, 
Chad M. Topaz$^{1,\ast}$
\\
\bf{1} Department of Mathematics, Statistics, and Computer Science, Macalester College, Saint Paul, Minnesota, United States of America
\\
\bf{2} Department of Mathematics, Harvey Mudd College, Claremont, California, United States of America
\\
$\ast$ E-mail: ctopaz@macalester.edu
\end{flushleft}

\section*{Abstract}

From bird flocks to fish schools and ungulate herds to insect swarms, social biological aggregations are found across the natural world. An ongoing challenge in the mathematical modeling of aggregations is to strengthen the connection between models and biological data by quantifying the rules that individuals follow. We model aggregation of the pea aphid, \emph{Acyrthosiphon pisum}. Specifically, we conduct experiments to track the motion of aphids walking in a featureless circular arena in order to deduce individual-level rules. We observe that each aphid transitions stochastically between a moving and a stationary state. Moving aphids follow a correlated random walk. The probabilities of motion state transitions, as well as the random walk parameters, depend strongly on distance to an aphid's nearest neighbor. For large nearest neighbor distances, when an aphid is essentially isolated, its motion is ballistic with aphids moving faster, turning less, and being less likely to stop. In contrast, for short nearest neighbor distances, aphids move more slowly, turn more, and are more likely to become stationary; this behavior constitutes an aggregation mechanism. From the experimental data, we estimate the state transition probabilities and correlated random walk parameters as a function of nearest neighbor distance. With the individual-level model established, we assess whether it reproduces the macroscopic patterns of movement at the group level. To do so, we consider three distributions, namely distance to nearest neighbor, angle to nearest neighbor, and percentage of population moving at any given time. For each of these three distributions, we compare our experimental data to the output of numerical simulations of our nearest neighbor model, and of a control model in which aphids do not interact socially. Our stochastic, social nearest neighbor model reproduces salient features of the experimental data that are not captured by the control.

\section*{Introduction}

From bird flocks to fish schools and ungulate herds to insect swarms, nature abounds with examples of animal aggregations \cite{ParHam1997,OkuLev2001,CamDenFra2001}. These groups may arise from environmental factors, social factors, or a combination of the two. Environmental factors induce organisms to move in relation to food sources, light sources, gravity, predators, wind, chemical gradients, and more. On the other hand, even in the absence of significant environmental cues, some animals aggregate because of their intrinsic social tendencies. Social forces such as attraction, repulsion and alignment occur when these organisms interact, sensing each other via sight, smell, hearing, and so forth \cite{Bre1954,MogEde1999,CouKraJam2002,EftVriLew2007,EftVriLew20070}. Social aggregations not only are examples of natural pattern formation, but on long time and space scales may influence disease transmission, food supply availability, ecological dynamics, and ultimately, evolution \cite{OkuGruEde2001,TilKar1998}. Additionally, the understanding of aggregations has been used to design algorithms in robotics, computer science, and engineering~\cite{BonDorThe1999,Pas2005}.

A central question in the study of aggregations pertains to the relationship between individual-level and group-level behaviors, and it is crucial to distinguish between these. Individual-level behaviors might include an organism's tendency to move closer to conspecifics, or to align its movement with that of its neighbors. Group-level properties describe characteristics of many individuals, such as the shape of an aggregation, its spatial density distribution, and its velocity distribution. The connection between individual and group-level behaviors is highly nontrivial, as is typical for a complex system \cite{ParEde1999}. One methodology for exploring this connection is through mathematical modeling. By constructing mathematical models that describe each individual organism's rules for movement, one can simulate and analyze the ensemble to investigate the aggregate behavior. Indeed, aggregation modeling is the subject of an intensive effort in the mathematical modeling community, explored in \cite{MogEde1999, FliGruLev1999,LevRapCoh2001,TopBer2004,TopBerLew2006,DOrChuBer2006,LevTopBer2009,BerTop2011,FetHuaKol2011,KolSunUmi2011,FetHua2012} and many dozens of other studies.

An ongoing challenge in aggregation modeling is to construct individual-level rules that are quantitatively accurate and well-tied to experimental data. Sometimes, modelers may attempt to calibrate models and infer parameters based on published field observations or experimental results, for example, as with recent studies of locust swarms \cite{TopBerLog2008,TopDorEde2012}. A more direct approach is to conduct experiments that track the motion of individuals and use the data, namely time series of organisms' positions and velocities, to construct models more directly. This approach has enhanced the understanding of fish schools \cite{TunKatIoa2013}, starling flocks \cite{BalCalCan2008}, and duck formations \cite{LukLiEde2010}. Presently, we consider social aggregation of the pea aphid, \emph{Acyrthosiphon pisum}. These particular aphids are significant both because they are severe crop pests \cite{EmdHar2007} and because they are a model organism in biology for studying disease transmission, insect-plant interactions, phenotypic plasticity, and more \cite{Int2010}.

Some foundational results on pea aphid movement appear in \cite{RoiMyeFra1979}. Aphids moving on the ground exhibit two dispersal behaviors: searching and running. In the searching behavior, aphids look for a nearby plant to inhabit. Running aphids, in contrast, travel far away from their original host plant, likely in an effort to evade predators. In\cite{RoiMyeFra1979}, aphids were exposed to predators while feeding on alfalfa plants. As a defense mechanism, aphids dropped from their feeding site and then traveled away from the original host plant. The average searching aphid made one turn every $6.67\ s$ and traveled $0.27\ cm/s$ while the average running aphid turned less frequently, every $27.8\ s$ and traveled faster, at $0.67\ cm/s$. In a given experimental run, aphids generally did not shift between the searching and running behaviors.

In the absence of predators, some aphids move infrequently \cite{RoiMyeFra1979}. When aphids are attacked by predators, the aphids employ defense mechanisms such as dropping from their location, running, or emitting a fluid droplet from the cornicle, a tube on the dorsal side of the last segment of the insect. The fluid droplet is composed of a mechanical protectant which temporarily paralyzes the jaws of the attacker \cite{Dix1958} and alerts nearby conspecifics and heterospecifics to the danger \cite{KisEdw1972}. The experiments in \cite{MonBaiSle2000} investigate the emission of this fluid droplet further by prodding aphids of various ages on the anterior portion of their thorax and recording the aphid as an emitter or non-emitter. Pre-reproductive aphids are the most likely age group to emit this fluid droplet, plausibly because they often live in close proximity to highly related kin. Once the aphids reach adulthood, it is more advantageous to invest energy in reproduction.

Despite the aforementioned account of chemical signaling, and while it is well-known that aphids aggregate around food sources \cite{WayCam1970}, much less is known about whether certain aphid species form aggregations that are intrinsically social. Aphid species \emph{Uroleucon nigrotuberculatum} and \emph{Uroleucon caligatum} experience lower mortality from generalist predators when aggregated \cite{Cap1987}, suggesting an evolutionary advantage for social aggregation. Other results on aphid aggregation appear in \cite{Kid1976a, Kid1976b, Str1967, Cap1987}. In \cite{Str1967}, pea aphids were placed in a chamber with five identical feeding stations. If the insects did not aggregate socially, one would expect an even distribution of aphids in each chamber, but this distribution was not observed. In both light and dark conditions, the aphids aggregated mainly in one or two of the feeding stations. Aphids in a dark environment still aggregated at statistically significant levels, albeit less strongly than in lit conditions, suggesting that vision may be one of the senses through which aggregation is activated. In contrast, in a key test in \cite{Str1967}, artificial aphids were placed behind the feeding stations such that their shadows were clearly visible. The aphids in the chamber were then allowed to choose one of the five feeding stations at random. In this test, the chamber with the aphid dummies did not show greater likelihood of being chosen, which implies that vision is not the only mechanism enabling social aggregation.

The experiments of \cite{Kid1976a, Kid1976b} found that lime aphids, \emph{Eucallipterus tiliae}, aggregate socially. Three studies in \cite{Kid1976b} are especially relevant. In the first study, an aphid was allowed to move and settle on a particular, uninhabited leaf. Its final position was marked and the aphid was removed. Trials were repeated on the same leaf with different individuals whose final positions were similarly marked. The distribution of settling locations was random, suggesting that microhabitats on a leaf do not influence aphids' movement. However, when multiple individuals were allowed to settle simultaneously on the leaf, they aggregated, suggesting that social interactions influence their movement. In the second study, between one and eleven aphids were already settled on a leaf, and one target aphid was placed on the leaf. When the target aphid approached a settled aphid (with approach defined as walking within $1\ cm$) on 82\% of the trials, the target aphid settled within $1\ cm$ of the other settled aphid. The third study examined aphid distribution for different population densities. In this study, each aphid had an associated virtual territory, defined as a circle of fixed radius around the insect, identical for all individuals. In experimental trials, the group was allowed to approach an equilibrium configuration. Then, the percent leaf coverage was computed as the area of the union of the territories divided by the area of the leaf. As the number of aphids was increased, the percent leaf coverage rose with decreasing slope, indicating close packing of the insects, ostensibly due to social interactions.

Given the evidence for social aggregation in some aphid species, our goal at present is to assess and model aggregation of the pea aphid. More specifically, in order to deduce individual-level rules, we conduct experiments to track the motion of aphids walking in a featureless circular arena. We observe that each aphid transitions stochastically between a moving and a stationary state. Moving aphids follow a correlated random walk. The probabilities of stopping and starting, as well as the random walk parameters, depend strongly on distance to an aphid's nearest neighbor. For large nearest neighbor distances, when an aphid is essentially isolated, its motion is ballistic. Aphids move faster, turn less, and are less likely to stop. In contrast, for short nearest neighbor distances, aphids move more slowly, turn more, and are more likely to become stationary; this behavior constitutes an aggregation mechanism. From the experimental data, we estimate the state transition probabilities and correlated random walk parameters as a function of nearest neighbor distance. With the individual-level model established, we assess whether it reproduces the macroscopic patterns of movement at the group level. To do so, we consider three distributions, namely distance to nearest neighbor, angle to nearest neighbor, and percentage of population moving at any given time. For each of these three distributions we compare our experimental data to the output of numerical simulations of our nearest neighbor model, and of a control model in which aphids do not interact socially. Our social nearest neighbor model reproduces salient features of the experimental data that are not captured by the control.

\section*{Experimental Methods}

To host aphid colonies, we grew fava bean plants, \emph{Vicia faba} (Johnny's Selected Seeds, Winslow, ME) with \mbox{6 - 7} seeds per pot in an approximately $20\ ^\circ C$ laboratory setting at 60\% - 70\% relative humidity. We stored plants in $45\ cm\ \times\ 45\ cm\ \times 45\ cm$ mesh enclosures (BugDorm, Taichun, Taiwan). Plants received $12\ hr$ of continuous light per day from a $120\ W$ grow lamp suspended $\text{5 - 7.5 }cm$ above the enclosure, or $\text{25 - 30 }cm$ above the plants. We considered plants to be mature enough to host aphids approximately two weeks after planting, when they reached a height of $15\ cm$ above the flower pot rim. We colonized each plant with one hundred pea aphids, \emph{A. pisum} (Nasco, Fort Atkinson, WI). We periodically cleaned enclosures when dirt or dead aphids accumulated. By seven days after colonization, plant health would deteriorate due to aphid feeding. At this point, we transferred the colony to fresh plants in a fresh enclosure. Aphids were then given several days to acclimate before being used in experimental trials.
	
We performed experiments on a vibration isolation table (IsoStation, Newport Corp., Irvine, CA) in a darkened lab in order to minimize effects of the ambient environment. The experimental arena consisted of a polypropylene circular ring, with a radius of $20\ cm$ and height of $3/16\ in$, enclosed between two $1/8\ in$ thick glass plates. We underlit the arena with a $24\ in \times24\ in$ LED light panel (AnythingDisplay, Nashua, NH) having a $6500\ ^\circ K$ pure white color temperature. In order to remove debris that might interfere with imaging, and to remove any biological material that might potentially be left from previous experimental runs, we cleaned the top and bottom glass plates with acetone, ethanol and compressed air before every trial. We lined the arena wall and ceiling with silicone oil to discourage aphids from occupying the arena's walls and ceiling.

Aphids are dimorphic insects that may develop into winged or wingless forms, depending on a complicated interaction between genetics and environment \cite{Bri2010}. Since we wished to track two-dimensional motion, and in order to minimize any behavioral variations due to age, we restricted our experimental trials to adult wingless aphids (as identified by sight). To initiate a trial, we selected individuals from a colonized plant, typically selecting a mix of aphids who appeared to be stationary and moving. Three trials incorporated 8, 10 and 18 aphids; the remaining six trials incorporated 27 - 35 aphids moving in the arena. We filmed the experiment using a 1080p high definition video camera (Sony Handycam HDR-SR12) placed $1.1\ m$ above the arena, with white balance calibrated to adjust for the effect of the light box as a background. After 45 minutes of filming we ceased recording and returned aphids to the colony.

\begin{figure}[!h]
\begin{center}
\includegraphics[width=\textwidth]{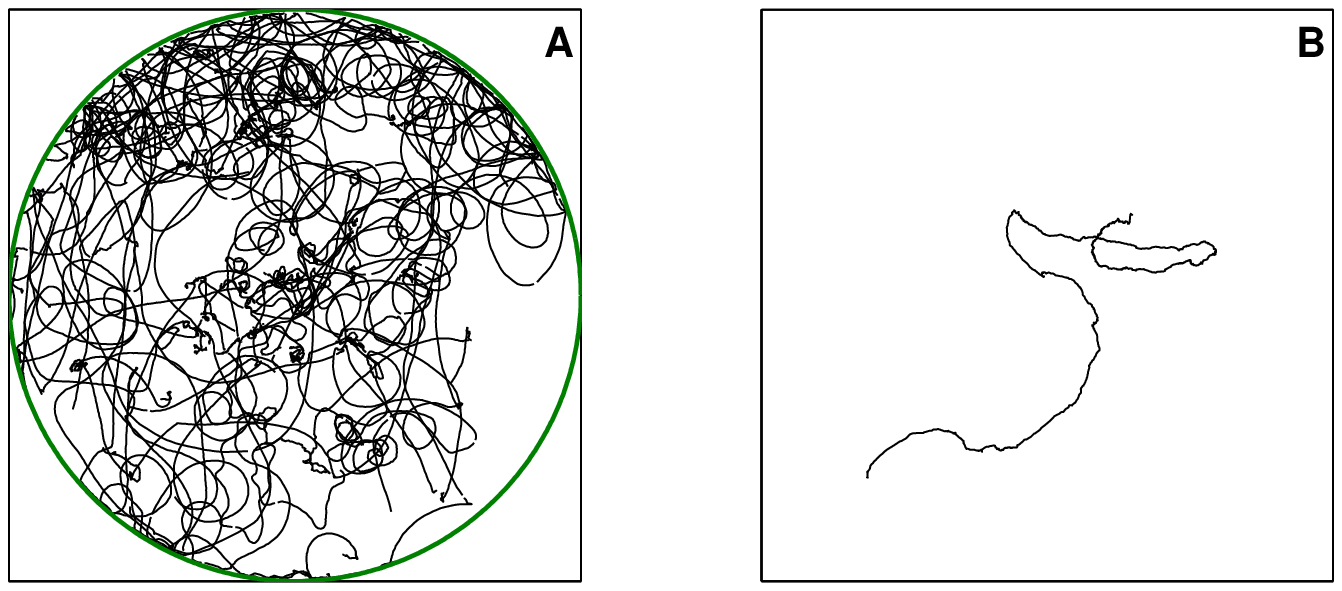}
\end{center}
\caption{
{\bf Visualizations of aphid movement in experiment.} (A) Trajectories of 28 aphids during approximately $15\ min$ of one experimental trial, as determined by motion tracking of video data. The green circle is the experimental arena with radius $20\ cm$. (B) Blow-up of a subset of a single aphid trajectory, shown in a $10\ cm \times 10\ cm$ zoom.}
\label{fig:sampletrajectories}
\end{figure}

To prepare our data for motion tracking, we converted raw video footage in .mts format to .mp4 using Handbrake video processing software with sampling in grayscale at $5\ fps$. We used QuickTime Pro to export the video into an image sequence of .tiff files, downsampled to 256 grays and $2\ fps$ to facilitate data processing. Using the ImageJ image processing package \cite{SchRasEli2012} we removed initial frames of each trial during which overhead lights were reflected, and cropped the rectangular video frames to a circular region corresponding to the experimental arena. We further processed images using \matlab's Image Processing Toolbox and the u-track 2.0 motion tracking package \cite{JaqLoeMet2008}. Specifically, we converted color images to black and white ones (to render the inside of the arena black) and denoised each frame. We ran u-track, which forms trajectories by linking identified aphid positions from frame to frame using a Kalman filter for motion propagation. The tracking process resulted in more trajectories than the number of aphids used in the trial due to the inherent difficulty of motion tracking. That is to say, a single aphid's track across the course of an experimental trial may be recognized as several, shorter trajectories by the tracking algorithm, but this does not affect our data analysis and modeling (more details appear in subsequent sections). Finally, we converted tracked aphid positions from pixel coordinates to real coordinates. Fig.\,~\ref{fig:sampletrajectories} shows examples of tracked data.

To prepare our raw data set for modeling (see next section) we enhanced it with several elementary, derived pieces of data, namely motion state (stationary or moving), step length (distance traveled in one frame), heading, turning angle, and distance to nearest neighbor. An aphid's step length in a current frame was calculated as magnitude of the difference between its current and previous positions. We considered an aphid to be moving in a given frame if its step length was sufficiently large. For small steps, corresponding to speeds less than $4 \times 10^{-2}\ cm/s$ (about 1/10 body length per second), we assumed the aphid to be stationary, with the small amount of movement attributed to noise in the video itself and errors in the aphid identification and tracking algorithms. An aphid's heading (the direction it was traveling in a given frame) was calculated by taking the angle of the difference between the aphid's current and previous position vectors. Finally, we calculated turning angle in a given frame as the difference in the current and previous heading. Our final data set consists of 1.2 million entries from the pooled data of nine experimental runs. Each entry contains an aphid's position, motion state, step length, heading, and turning angle.

\section*{Mathematical Modeling of Individual-Level Behaviors}

Based on the observation that aphids in the experimental trials transitioned between stationary and moving behavior, we propose a probabilistic two-state model to describe aphid movement and social interaction dynamics. Let $\prob{\ms}$ represent the probability that a moving aphid in a given frame transitions to a stationary state in the next frame. Similarly, let $\prob{\sm}$ represent the probability that a stationary aphid in a given frame transitions to a moving state. Perhaps the simplest model that accounts for social interactions allows these probabilities to depend solely on the distance to an aphid's nearest neighbor, $\dnn$. The underlying biological assumptions leading to this model are that aphids sense isotropically (perhaps due to a combination of visual, auditory, and olfactory inputs), that they are affected by the minimum possible social information, and that they do not react to the speed and orientation of their neighbor. We will show that this minimal model reproduces certain salient features of the experimental data.

Moving aphids appear (naively) to follow a correlated random walk \cite{Tur1998}; see Fig.\,\ref{fig:sampletrajectories}B. In an (unbiased) correlated random walk, an individual walks in a straight line of a certain (random) step length $\step$, turns from its previous heading at an angle $\turn$ that is random but drawn from a mean-zero distribution, and then repeats. In our model, we will assume that the correlated random walk parameters depend solely on distance to nearest neighbor, similar to the transition probabilities discussed above. For step length, we choose the simplest model, meaning that there is no spread in the step length distribution. A moving aphid's step length $\step$ depends deterministically on its distance to nearest neighbor $\dnn$. For turning angle $\turn$, the mean of the distribution is zero by the assumption of symmetry of the correlated random walk. Therefore, we model dependence on $\dnn$ in the spread $\turnspread$ of the turning angle distribution.

We will now quantify our four model parameters: probability of a moving aphid stopping ($\prob{\ms}$), the probability of a stationary aphid starting to move ($\prob{\sm}$), a moving aphid's step length traveled in one frame ($\step$), and the spread of the turning angle distribution that a moving aphid obeys ($\turnspread$). Each of these will depend on distance to an aphid's nearest neighbor $\dnn$ through simple functional forms with three or four parameters, which we estimate from experimental data below.

\begin{figure}[!h]
\begin{center}
\includegraphics[width=\textwidth]{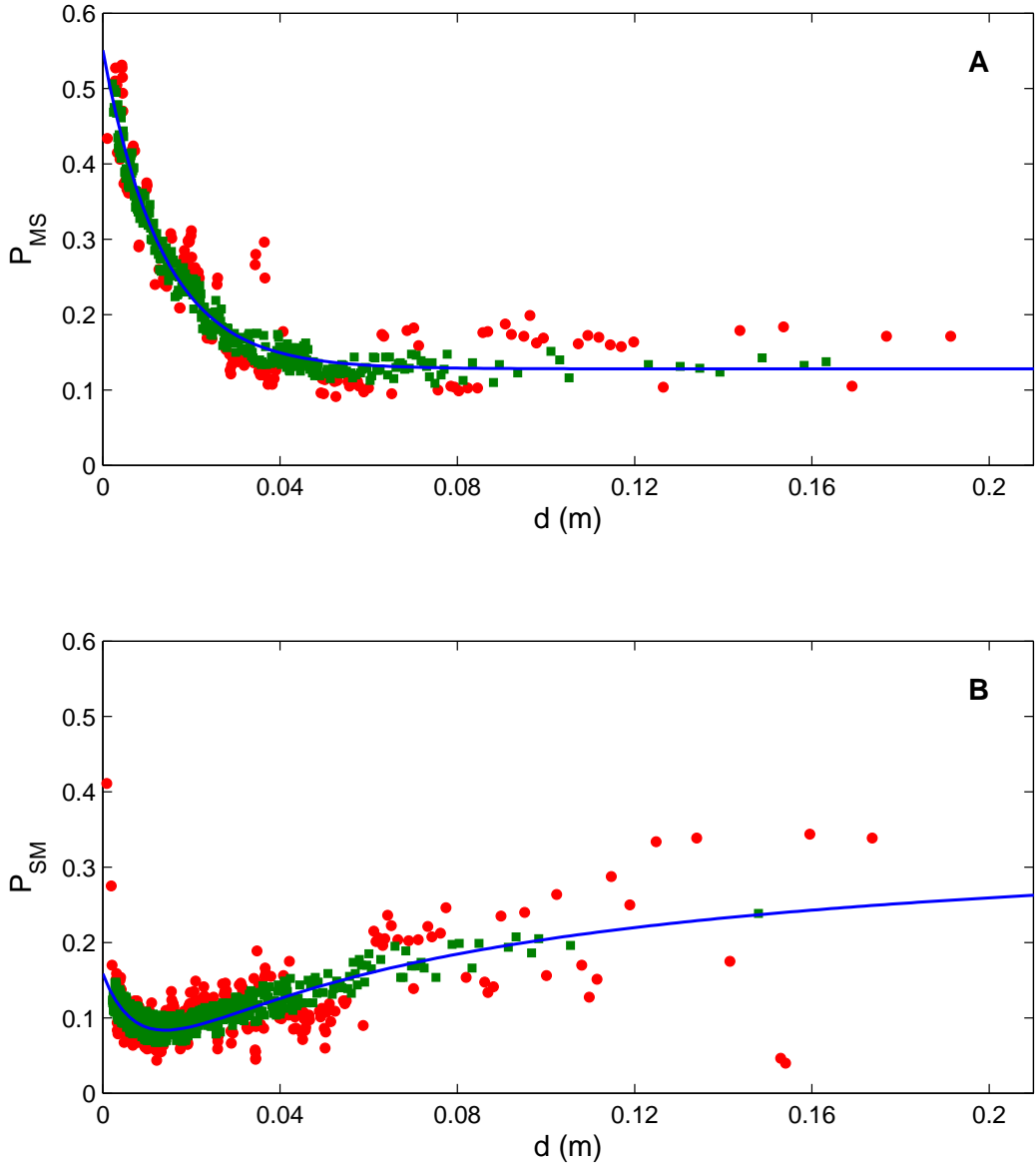}
\end{center}
\caption{
{\bf State transition probabilities $\boldsymbol{\prob{\ms}}$ and $\boldsymbol{\prob{\sm}}$ as a function of distance to an aphid's nearest neighbor, $\boldsymbol{\dnn}$ (in $\boldsymbol{m}$).} (A) $\prob{\ms}$, the probability that an aphid moving in a given timestep becomes stationary at the next timestep. Each data point represents the probability within a bin of $800$ elements from our experimental data set, where the data are binned by $\dnn$. The probability is calculated via a simple frequency count according to Eq.\,\eqref{eq:frequencycount}. The overall dependence of the data on $\dnn$ is modeled with Eq.\,\eqref{eq:pmsfunctionalform}, which describes an increased probability of an aphid settling if a neighbor is nearby. Best fit parameters appear in the text; the coefficient of determination is $R^2=0.92$. To give a further sense of the efficacy of the fit, we display each point according to the standard error of the mean within the bin it represents. If the model curve passes within two standard errors of the estimated value, we show it as a green square; otherwise, it is a red dot. (B) Like (A), but for the probability $\prob{\sm}$ that a stationary aphid starts moving. The model is Eq.\,\eqref{eq:psmfunctionalform}, describing higher aphid mobility at very short and very long $\dnn$. Here, $R^2=0.52$; see text for discussion.}
\label{fig:transitions}
\end{figure}

To estimate the transition probabilities $\prob{\ms}$ and $\prob{\sm}$, we note that our data set (see previous section) includes a motion state for each entry. We can classify every transition that occurs in the data set as stationary to stationary ($\sstat$), stationary to moving ($\sm$), moving to moving ($\mm$) or moving to stationary ($\ms$). We divide the data set in two, with $\sstat$ and $\sm$ in one subset and $\mm$ and $\ms$ in the other. For each subset, we generate bins of 800 data points where binning is performed according to $\dnn$. Within each bin, we estimate the probability of a transition as the ratio of the number of occurrences of the transition to the total number of observations. For instance, within a given bin, we estimate $\prob{\ms}$ as
\begin{equation}
\label{eq:frequencycount}
\prob{\ms} = \frac{\ms\text{ occurrences}}{\ms\text{ occurrences} + \mm\text{ occurrences}}.
\end{equation}
We then form a scatterplot of the probability within each bin versus the midpoint of the bin, resulting in Fig.\,\ref{fig:transitions}.

The probability $\prob{\ms}$, shown in Fig.\,\ref{fig:transitions}A, appears to decrease monotonically with $d$ and level off. We model this decrease with the functional form,
\begin{equation}
\prob{\ms}(\dnn) = \prob{\ms}^\infty + \left(\prob{\ms}^0 - \prob{\ms}^\infty\right) \exp{-\dnn/d_{MS}}.
\label{eq:pmsfunctionalform}
\end{equation}
The probability $\prob{\ms}^0$ represents the probability that an aphid will become stationary when infinitesimally close to its nearest neighbor, whereas $\prob{\ms}^\infty$ is the probability of transitioning when isolated, that is, even in the absence of sensed neighbors. The length scale $d_{MS}$ characterizes the transition between the two limiting regimes of $\dnn$. The choice of a decaying exponential function not only agrees well with the data (as discussed presently) but has biological motivation. If one assumes that the motion state transition occurs due to sensing, and that the sensory input an aphid receives has a constant probability of failure per distance displaced from its source, then one obtains an exponential model, a common choice for aggregation modeling \cite{MogEdeBen2003}. Overall, the model Eq.\,\eqref{eq:pmsfunctionalform} reflects aphids being more likely to settle near other individuals, in order to aggregate.

To fit Eq.\,\eqref{eq:pmsfunctionalform} to the experimental data, we first observe that $\prob{\ms}^\infty$ and $\prob{\ms}^0$ appear linearly while $d_{MS}$ appears nonlinearly. We minimize the root-mean-square (RMS) error of the fit by scanning across values of $d_{MS}$ and at each value, performing a least squares fit for the two linear parameters. We find $\prob{\ms}^\infty \approx 0.1280$, $\prob{\ms}^0 \approx 0.5508$, and $d_{MS} \approx 0.0134\ m$, resulting in a fit (shown as the blue curve) with a high coefficient of determination, $R^2 = 0.92$. To give a further sense of the efficacy of the fit, it is helpful to consider the standard error in each bin, which is given by
\begin{equation}
\sqrt{\frac{\prob{\ms}(1-\prob{\ms})}{N}},
\end{equation}
where $N$ is the number of aphids per bin and $\prob{\ms} = \prob{\ms}(d)$ is the probability of transition within the bin. Green squares (red dots) represent bins for which the corresponding model prediction is within (outside of) two standard errors of the estimated $\prob{\ms}$.

The probability $\prob{\sm}$ is shown in Fig.\,\ref{fig:transitions}B. Unlike $\prob{\ms}$ which decreases monotonically, $\prob{\sm}$ has a minimum at short distances. We choose the functional form
\begin{equation}
\label{eq:psmfunctionalform}
\prob{\sm}(\dnn)= \prob{\sm}^0 \exp{-\dnn/d_{SM}} + \prob{\sm}^\infty \frac{\dnn}{\dnn+\Delta_{SM}}.
\end{equation}
The first (exponential) term is repulsive, consistent with the notion that aphids avoid settling too close to others. The second (rational) term is attractive, modeling the tendency of solitary aphids to move in order to aggregate. We fit this functional form to the data through a procedure similar to $\prob{\ms}$, except that we must now search over a grid of two nonlinear parameters, $d_{SM}$ and $\Delta_{SM}$. We find $\prob{\sm}^0 \approx 0.1587$, $\prob{\sm}^\infty \approx 0.3552$, $d_{SM} \approx 0.0079\ m$, and $\Delta_{SM} \approx 0.0739\ m$. To compare each data point to the model, we use the same green square/red circle scheme as above. The overall fit has $R^2 = 0.53$. This coefficient of determination, substantially lower than for $\prob{\ms}$, is likely due to the large scatter of the data for large $\dnn$, which may reflect two sources of error. First, imaging and tracking of aphids is more difficult when they are in the vicinity of the boundary of the arena, and aphids at large $\dnn$ are more likely to be near a boundary. Second, it is possible that there is an explicit effect of the boundary on aphids' behavior which we have not modeled here.

We tried several functional forms (including linear combinations of exponentials) but choose Eq.\,\eqref{eq:psmfunctionalform}, which minimizes the RMS error with two pairs of parameters. We believe the exact functional form is less important than the trends of higher mobility at both very short and very long distances.

\begin{figure}[!h]
\begin{center}
\includegraphics[width=\textwidth]{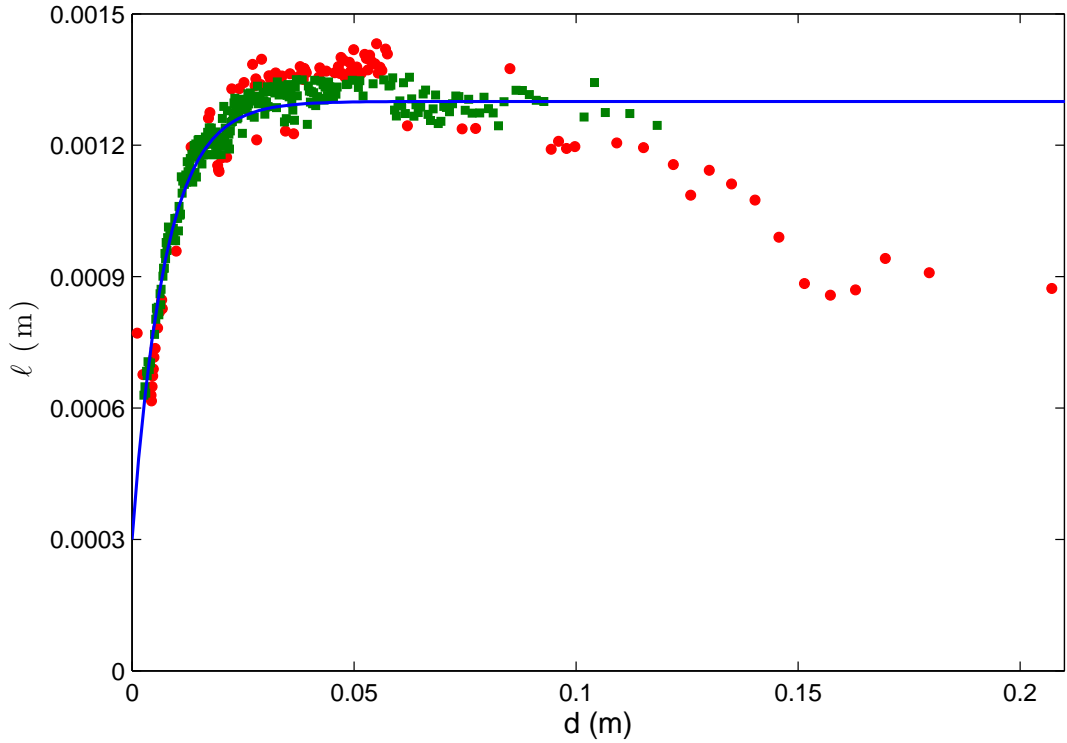}
\end{center}
\caption{
{\bf Correlated random walk step length $\boldsymbol{\ell}$ (in $\boldsymbol{m}$) per frame as a function of distance to an aphid's nearest neighbor $\boldsymbol{\dnn}$ (in $\boldsymbol{m}$).} Each data point represents the mean step length within a bin of $800$ elements from our experimental data set, where the data are binned by $\dnn$. The overall dependence of the data on $\dnn$ is modeled with Eq.\,\eqref{eq:stepfunctionalform}, which captures the tendency of aphids to aggregate simply by traveling less when in the vicinity of others. Best fit parameters appear in the text; the coefficient of determination is $R^2=0.82$. To give a further sense of the efficacy of the fit, we display data points according to the same scheme used in Fig.\,\ref{fig:transitions}. Green squares (red dots) represent data bins for which the model prediction falls within (outside) two standard errors of the experimental mean.}
\label{fig:steplength}
\end{figure}

We now turn to the parameters governing moving aphids' correlated random walks. Fig.\,\ref{fig:steplength} shows the mean step length as a function of $\dnn$, with each point in the scatterplot corresponding to a bin of 800 data points. Because there is a coherent rise in the data for small $\dnn$, we consider the model
\begin{equation}
\step(\dnn) = \step^\infty + \left(\step^0 - \step^\infty\right) \exp{-\dnn/d_{\step}}.
\label{eq:stepfunctionalform}
\end{equation}
According to this model, aphids with neighbors nearby take short steps, and the step length increases and saturates as $\dnn$ increases. Using a similar fitting procedure to $\prob{\ms}$ and $\prob{\sm}$, we find $\step^\infty \approx 0.0013\ m$, $\step^0 \approx 0.0003\ m$, and $d_{\step} \approx 0.0074\ m$. Within each bin, the standard error around the mean is $s/\sqrt{N}$ where $N$ is the number of observations and $s$ is the sample standard deviation. To compare experimental bins with the model prediction, we use the same green squares/red dot visualization as above. For the overall fit, we find $R^2=0.82$. The data decrease moderately from our model curve for $\dnn > 0.1\ m$, which is half the radius of our experimental arena. Once again, we believe that we may be seeing biases due to the boundary and the increased difficulty of motion tracking near the boundary.

\begin{figure}[!ht]
\begin{center}
\includegraphics[width=0.9\textwidth]{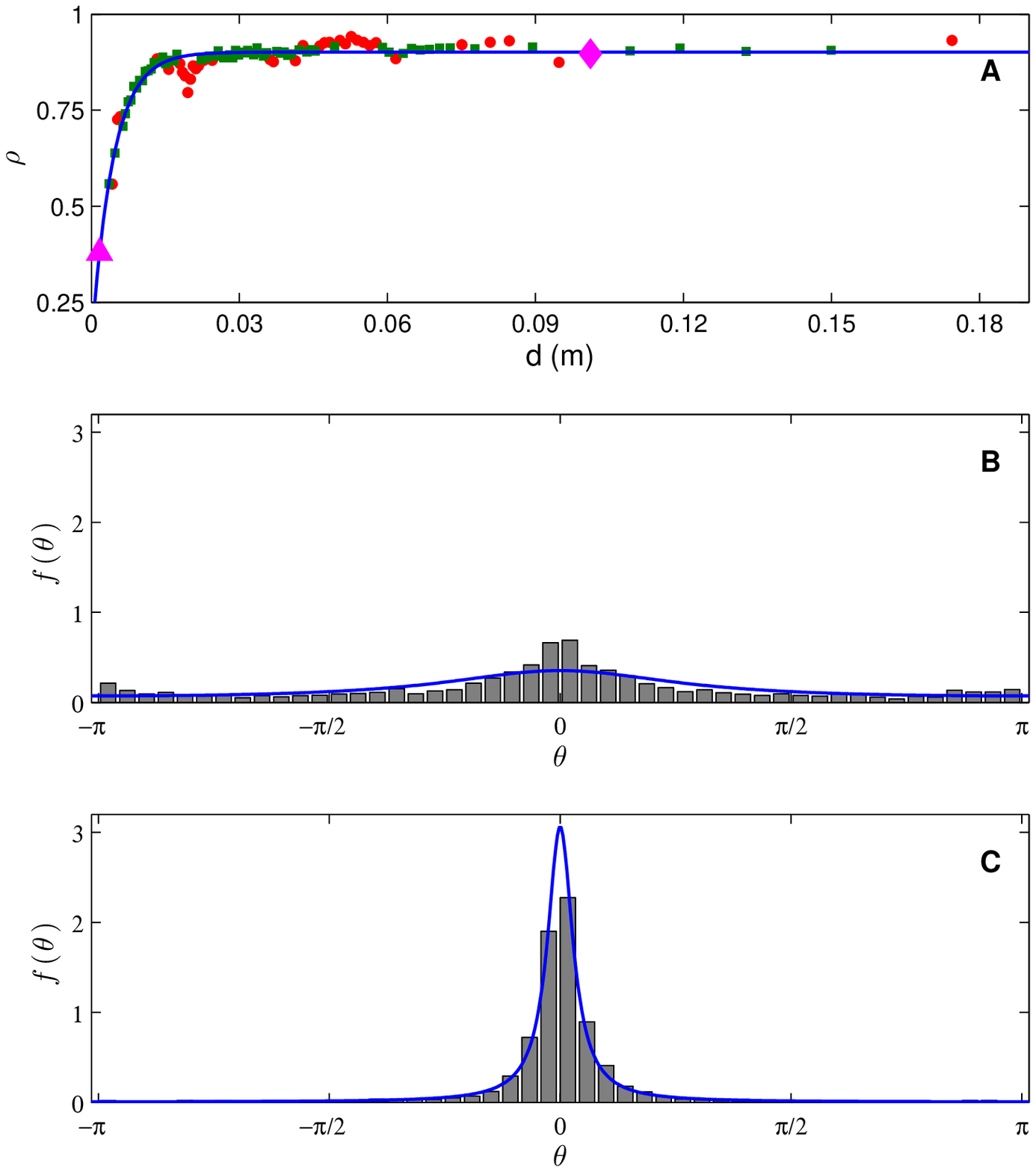}
\end{center}
\caption{
{\bf Correlated random walk turning angle $\boldsymbol{\turn}$.} (A) Turning angle distribution parameter $ 0 < \turnspread < 1$ as a function of distance to an aphid's nearest neighbor. Here, $\turnspread$ is a parameter in the zero-mean wrapped Cauchy distribution Eq.\,\eqref{eq:wrappedcauchy} used to model turning angle $\turn$. Each data point represents the experimentally measured value of $\turnspread$ within a bin of $2400$ elements from our experimental data set, where the data are binned by $\dnn$. The overall dependence of the data on $\dnn$ is modeled with Eq.\,\eqref{eq:turnspreadfunctionalform}, which captures the tendency of aphids to aggregate by taking wider turns when in the vicinity of others, leading to motion that is more diffusive and less ballistic. Best fit parameters appear in the text; the coefficient of determination is $R^2=0.99$. Green circles (red dots) points represent data bins for which the model prediction falls within (outside) a 95\% confidence interval around the experimentally measured $\rho$, where the interval is constructed by resampling our original data $20,000$ times. (B) Normalized histogram showing the experimental turning angle distribution within the data bin corresponding to the magenta triangle in (A). The blue curve shows the wrapped Cauchy distribution predicted by our model. (C) Like (B), but for the magenta diamond.}
\label{fig:turningangle}
\end{figure}

Finally, we model the spread of the distribution of turning angles $\turn$. We bin $\turn$ values by $\dnn$ with 2400 values per bin (larger than the previously used value of 800 in order to help reduce the standard error within each bin). As alluded previously, within every data bin, the distribution is strongly peaked around zero; see the examples in Fig.\,\ref{fig:turningangle}B and Fig.\,\ref{fig:turningangle}C. Therefore, to capture the effect of neighbors, it is necessary to model the spread of the distribution of $\turn$, which indeed appears to depend on $\dnn$. Since $\turn$ is an angular distribution, it exists on the interval $[-\pi,\pi]$. Wrapped normal distributions give a poor fit to our data (not shown). We instead select the wrapped Cauchy distribution \cite{Fis1993} centered at zero,
\begin{equation}
\label{eq:wrappedcauchy}
f(\turn)=\frac{1}{2\pi}\frac{1-\turnspread^{2}}{1+\turnspread^{2}-2\turnspread\cos \theta},
\end{equation}
where $0<\turnspread<1$ is a parameter governing the spread of the distribution. Small values of $\turnspread$ correspond to more spread distributions, whereas values closer to one result in strongly peaked distributions. Fig.\,\ref{fig:turningangle}A shows $\turnspread$ as a function of $\dnn$ for the binned data. As a model, we select the functional form
\begin{equation}
\turnspread(\dnn) = \turnspread^\infty + \left(\turnspread^0 - \turnspread^\infty\right) \exp{-\dnn/d_{\turnspread}}.
\label{eq:turnspreadfunctionalform}
\end{equation}
According to this model, aphids with nearby neighbors will turn more often at wider angles, resulting in motion that is less ballistic and more diffusive.

Fitting the model as described previously, we find $\turnspread^\infty \approx 0.9013$, $\turnspread^0 \approx 0.1387$, and $d_{\turnspread} \approx 0.0044\ m$. To compare the experimental data and the model within a given bin, we calculate a $95\%$ confidence interval by resampling the data in each bin thousands of times, calculating $\turnspread$ each time, and considering the resulting distribution of values of $\turnspread$. If the value of $\turnspread$ predicted by our model falls within the central 95\% of the sampling distribution, we show the data point in Fig.\,\ref{fig:turningangle}A as a green square; otherwise it is a red dot. For the fit of Eq.\,\eqref{eq:turnspreadfunctionalform}, we find $R^2=0.99$.

In summary, our model consists of just four quantities: $\prob{\ms}$, $\prob{\sm}$, $\step$, and $\turnspread$. Each of these depends on $\dnn$ via three or four parameters. In total, we have fit 13 parameters, but we note that there are over one million entries in our data set.

\begin{figure}[!ht]
\begin{center}
\includegraphics[width=\textwidth]{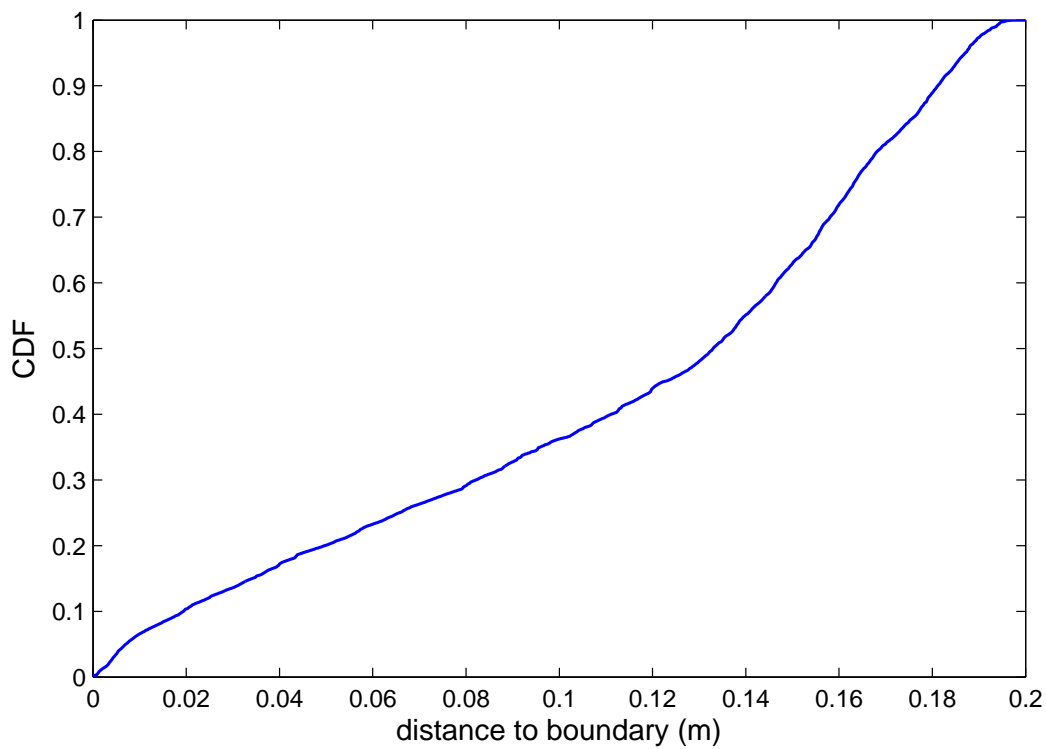}
\end{center}
\caption{
{\bf Cumulative distribution of aphids as a function of distance to arena boundary (in $\boldsymbol{m}$) for experimental data set.} The circular experimental arena has a radius of $0.2\ m$. Only 10\% of the data set corresponds to aphids within $2\ cm$ (about five body lengths) of the boundary.}
\label{fig:boundary}
\end{figure}

As alluded previously, one component ignored in the model is the arena's boundary. While it is quite likely that the presence of a boundary wall influences aphids' movement, the majority of our data set is composed of aphids far from the boundary. Fig.\,\ref{fig:boundary} shows the cumulative distribution function of distance to boundary for the entire data set. Only 10\% of our data is within $2\ cm$ of the boundary (4 or 5 aphid body lengths), and we leave the quantification of boundary effects as future work.

With our model for individual-level behavior established, we will presently assess the degree to which it reproduces group-level behaviors. For comparison and contrast, we also consider a control model in which aphids do not interact at all. For this non-interaction model, we use the asymptotic (limit of large $\dnn$) values of the parameters in our individual-level model. That is, we set  $\prob{\ms}=\prob{\ms}^\infty$, $\prob{\sm}=\prob{\sm}^\infty$, $\step=\step^\infty$, and $\turnspread=\turnspread^\infty$.

\section*{Simulation and Analysis of Group-Level Behaviors}

We now shift our focus to group-level behaviors. We compare the experimental data ($\experiment$) with data simulated from the two models developed above, namely the one in which aphids interact with their nearest neighbor (model $\inter$) and the one in which aphids do not interact (model $\nonint$). For each model, we carry out simulations parallel to each experimental run, that is, having the same initial aphid positions and containing the same number of frames. We augment the individual-level behaviors with a rule for what simulated aphids do if they encounter the (simulated) arena boundary. If an aphid travels to a new position that would be outside of the arena, we apply a simplistic reflective boundary condition in which the angle of incidence on the boundary equals the angle of reflection. Also, we let the distance the aphid travels once it reflects off the wall be the distance it would have travelled beyond the boundary.

We will compare three different group-level behaviors by studying their corresponding cumulative distribution functions as computed across each data set. It will be convenient to call these $F_i^{\experiment}$, $F_i^\inter$, $F_i^\nonint$, where the subscript $i$ indexes the distribution (since it is discrete). Our strategy will be to make three pairwise comparisons for each group-level behavior, namely $F_i^{\experiment}$ vs. $F_i^\inter$, $F_i^{\experiment}$ vs. $F_i^\nonint$, and $F_i^\inter$ vs. $F_i^\nonint$. It is also convenient to define the underlying probability distributions, $f_i^{\experiment}$, $f_i^\inter$, $f_i^\nonint$. For each pairwise comparison we will calculate several different quantities. A simple comparison is the distance between median values of the probability distributions, which we refer to as $\mediandiff$. Another choice is the Kolmogorov-Smirnov distance $D_{KS}$ \cite{Kol1933,Smi1948}, a common nonparametric measure equal to the maximum vertical distance between two cumulative distributions. Finally, we consider the Kullback-Leibler divergence $D_{KL}$ \cite{KulLei1951}. This quantity measures the information lost when a distribution $f_i^2$ is used to approximate another distribution, $f_i^1$. It is defined as
\begin{equation}
D_{KL} \left(f_i^1 || f_i^2\right)= \sum_i \ln \left( \frac{f_i^1}{f_i^2}\right) f_i^1,
\end{equation}
where for us, the superscript $1$ and $2$ will refer to one of our three data sets. Results appear in \mbox{Tables~\ref{tab:statsdnn} - \ref{tab:statspercent}}. We do not perform statistical hypothesis testing using $\mediandiff$, $D_{KS}$, and $D_{KL}$ because we have no null hypothesis that our models and experiment produce statistically indistinguishable data. Rather, we expect that they are different, and we simply use empirical measures to assess the closeness of the model distributions to the experimental one.

\begin{figure}[!ht]
\begin{center}
\includegraphics[width=\textwidth]{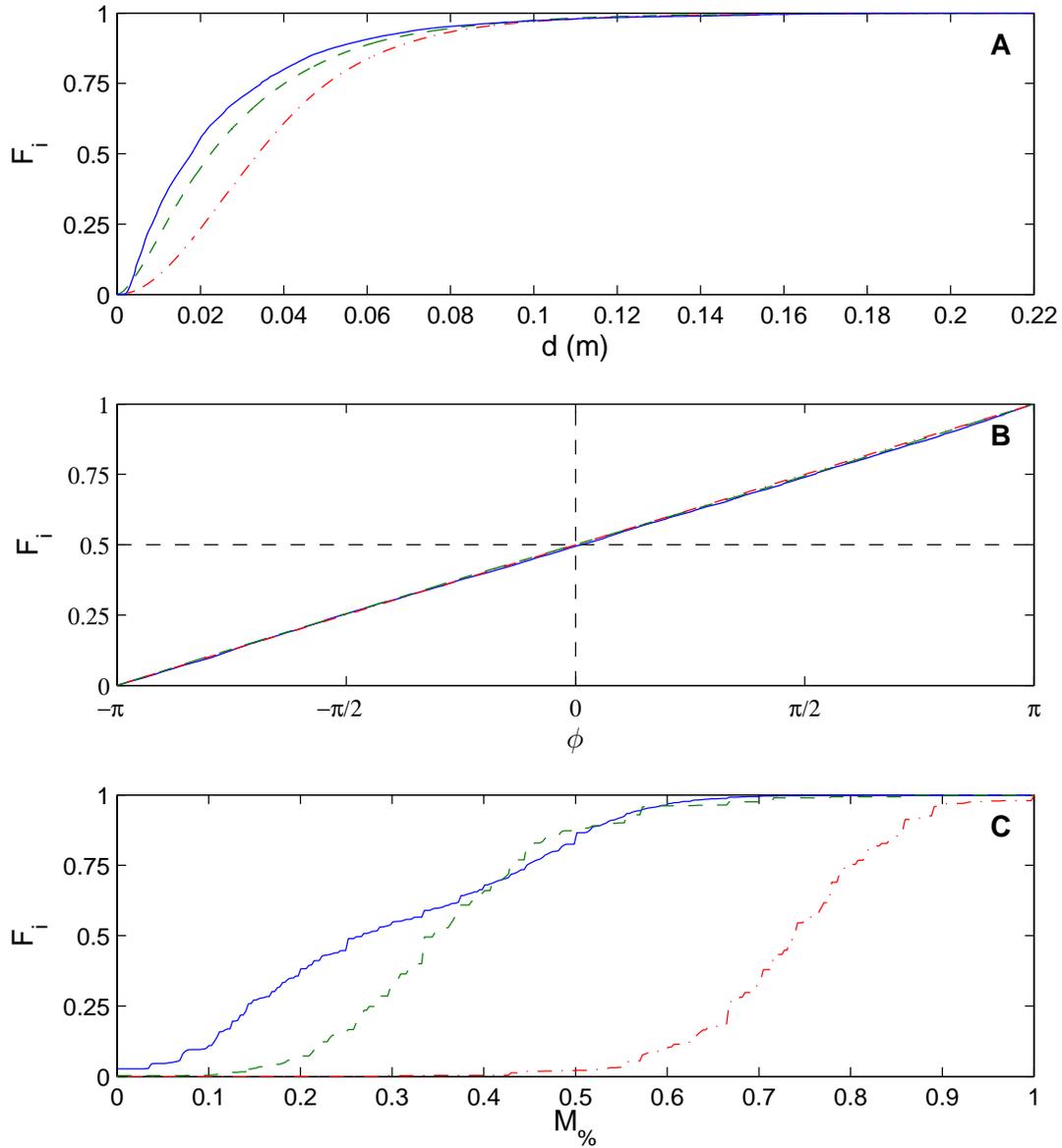}
\end{center}
\caption{
{\bf Pea aphid group-level behaviors in experiment, a social interaction model, and a control (non-interacting) model.} (A) Cumulative distributions $F_i$ of distance to nearest neighbor $\dnn$ (in $m$) for experimental data set (solid blue), social interaction model (dashed green), and non-interacting model (dot-dashed red). (B) Like (A), but the cumulated quantity is angle to nearest neighbor $\ann$ (relative to an aphid's heading $\dir$). (C) Like (A), but the cumulated quantity is $M_\%$, fraction of the aphid population moving in a given frame. As compared to the curves in (A) and (B), the more staircase-like appearance of these curves arises simply from the fact that the variable being cumulated is discrete (percentage of aphids in a group of several dozen) as opposed to the continuous variables in (A) and (B). For (A) - (C), measures of the difference between the distributions are given in Tables \ref{tab:statsdnn} - \ref{tab:statspercent} respectively.}
\label{fig:group}
\end{figure}

The first group-level behavior we consider is the distribution of nearest neighbor distances $d$ that emerges through an experiment or simulation. The cumulative distributions are shown in Fig.\,\ref{fig:group}(A), with $\experiment$ as solid blue, $\inter$ as dashed green, and $\nonint$ as dot-dashed red. Statistical measures are given in Table \ref{tab:statsdnn}. We see that $\mediandiff$ is smaller for $\experiment$ vs. $\inter$ than for $\experiment$ vs. $\nonint$ by approximately a factor of two. Put differently, the shorter median $\dnn$ for $\inter$ (as opposed to $\nonint$) indicates that the social behaviors in the model indeed promote aggregation. The experimental curve has an even shorter $\dnnmedian$. Model $\inter$ appears to capture some (but not all) of the aggregative tendency seen in the experiment. The Kolmogorov-Smirnov distance, $D_{KS}$, is smaller between $\experiment$ and $\inter$ than $\experiment$ and $\nonint$, as is $D_{KL}$. Thus, by all three measures, $\inter$ captures more of the experimental behavior than $\nonint$ does.

The second group-level behavior we consider is the distribution of angle to nearest neighbor, $\ann$, measured relative to an aphid's heading $\dir$. The cumulative distributions and statistical information appear in Fig.\,\ref{fig:group}(B) and Table \ref{tab:statsann}. The graph reveals that $\experiment$, $\inter$, and $\nonint$ all give rise to a uniform distribution of relative orientation (reflected by the linear cumulative profile). Therefore, aphids in experiment and in both models do not preferentially align towards their nearest neighbors. 

Finally, we consider the third group-level behavior, the distribution of the fraction $\fracmoving$ of aphids moving at a given time. The cumulative distributions and statistical information appear in Fig.\,\ref{fig:group}(C) and Table \ref{tab:statspercent}. They are strikingly different. As with the distributions for $d$, $\inter$ reproduces much more of the behavior of $\experiment$ than $\nonint$ does. The extreme rightward shift of the red curve indicates that the mobility of aphids is much higher in $\nonint$; put differently, aphids in this model do not aggregate and settle nearly as much as in $\experiment$ and $\inter$.

\section*{Conclusion}

Through experiment and modeling, we have investigated the movement, social behavior, and aggregation of the pea aphid. Motion-tracked experimental data gives rise to a two-state model in which aphids transition stochastically between stationary and moving states. Moving aphids follow a correlated random walk. The state transition probabilities $\prob{\ms}$ and $\prob{\sm}$, the random walk step length $\step$, and the random walk turning angle distribution spread $\turnspread$ all depend on distance to an aphid's nearest neighbor, $d$. These four quantities have each been fit with a functional form incorporating three or four parameters whose values we estimated from the experimental data. To assess the efficacy of our model in reproducing group-level behaviors, we compared experimental data to outputs of our social nearest neighbor model and a control (noninteracting) model. We found that the social model reproduces the distribution of nearest neighbors and the distribution of fraction of moving aphids better than the control model. The experiment and both models display no difference for a third group-level property, namely angle to nearest neighbor.

Our mathematical model is strikingly different from some previous data-driven aggregation models. The model of golden shiner fish in~\cite{TunKatIoa2013} and the model of surf scoter ducks in \cite{LukLiEde2010} are primarily deterministic, describing organisms that simultaneously attract, repel, and align. In these studies, noise additively modulates an organism's intended direction at each time step, presumably to describe errors in sensing and movement capabilities. In contrast, our model has rules that are fundamentally random. Fig.\,~\ref{fig:transitions} shows that aphids under similar conditions (same distance to nearest neighbor) display different behaviors (transitioning vs. not transitioning motion state). Fig.\,~\ref{fig:steplength} and Fig.\,~\ref{fig:turningangle} suggest that the movement process for aphids is a random walk.

The biological conclusions of our work are as follows. First, we have provided strong quantitative evidence that pea aphids display social behavior, in that an individual's movement in a featureless environment is influenced by its nearest neighbor.

Second, we have gained insight into the mechanism by which aphids aggregate. The probability of a stationary aphid starting to move decreases if a neighbor is nearby. The probability of a moving aphid stopping increases if a neighbor is nearby. These two behaviors promote aggregation. Further, aphids that are moving take shorter steps and turn more when in the vicinity of neighbors, promoting motion that is more diffusive and less ballistic (that is, less likely to move it away from the neighbor). This is reminiscent of the classic run-and-tumble model of bacteria \cite{BerBro1972}. In short, aggregation occurs through movement decreasing in the proximity of other aphids as opposed to direct locomotion towards individuals or clusters.

Finally, our model of individual-level behavior gives some feeling for the sensing range of the aphid. We recall the exponential length scales $\dnn_{\ms} \approx 0.0134\ m$, $\dnn_{\sm} \approx 0.0079\ m$, $\dnn_\step \approx 0.0074\ m$, and $\dnn_\turnspread \approx 0.0044\ m$. These characteristic length scales are on the order of 1 - 3 aphid body lengths.

As evidenced by the metrics in the previous section, our individual-based social model reproduces group-level behaviors much better than a control model. Nonetheless, we have not captured all of the experimental complexity in our simple model. As mentioned throughout, we have ignored the effects of the boundary. Further work could attempt to quantify more precisely the rules an aphid obeys when it encounters an immovable obstacle such as a boundary. Additionally, our model is arguably the simplest possible social model, in which social effects depend on a single nearest neighbor. One could investigate the degree to which an aphid responds simultaneously to multiple neighbors, keeping in mind the limits of aphid cognition. Finally, it could be interesting to augment our work, which describes aphid aggregation the absence of environmental cues, with a consideration of external factors such as nutrition sources. Such an investigation might shed further light on the aphid's role as a destructive crop pest.

\section*{Acknowledgments}
Ken Moffett of the Macalester College machine shop built the experimental arena. Matthew Beckman of Augsburg College provided advice on our experimental setup. Raibatak Das of the University of Colorado, Denver shared a template of \matlab\ code helpful in our image processing and tracking. We benefitted from statistical discussions with Alicia Johnson, Victor Addona, and Danny Kaplan. As part of his undergraduate research experience at Macalester College, Trevor McCalmont contributed to a prototype of the experiment and model during early stages of this work. We are grateful to Macalester College for laboratory space in the XMAC (eXperiment, Modeling, Analysis and Computation) laboratory.

\bibliography{master_bibliography}

\section*{Figure Legends}

\section*{Tables}

\begin{table}[!ht]
\caption{
\bf Measures comparing cumulative distributions of distance to nearest neighbor $\boldsymbol{\dnn}$ in experiment ($\boldsymbol{\experiment}$), a social interaction model ($\boldsymbol{\inter}$) and a noninteracting control model ($\boldsymbol{\nonint}$).}
\begin{tabular}{lccc}
\hline
{\bf Comparison} & $\mathbf{\mediandiff}$ & $\mathbf{D_{KS}}$ & $\mathbf{D_{KL}}$\\
\hline
$\experiment$ vs. $\inter$ & 0.0046 & 0.1083 & 0.0835 \\
$\experiment$ vs. $\nonint$ &  0.0159 & 0.3226 & 0.3873 \\
$\inter$ vs. $\nonint$ & 0.0113 & 0.2181 & 0.1668 \\
\hline
\end{tabular}
\begin{flushleft}By measures of the difference between median values $\mediandiff$ (in $m$), the Kolmogorov-Smirnov distance $D_{KS}$, and the Kullback-Leibler divergence $D_{KL}$, the cumulative distribution of $\inter$ comes closer to $\experiment$ than $\nonint$ does. Since $\mediandiff$ is a dimensioned quantity, it is meaningful to compare values to an aphid body length, approximately $0.004\ m$. $\experiment$ and $\inter$ have median values that differ by a body length, while the other two comparison have median differences an order of magnitude larger. The three distributions are shown in Fig.\,\ref{fig:group}(A).
\end{flushleft}
\label{tab:statsdnn}
 \end{table}

\begin{table}[!ht]
\caption{
\bf Measures comparing cumulative distributions of angle to nearest neighbor $\boldsymbol{\ann}$ in experiment ($\boldsymbol{\experiment}$), a social interaction model ($\boldsymbol{\inter}$) and a noninteracting control model ($\boldsymbol{\nonint}$).}
\begin{tabular}{lccc}
\hline
{\bf Comparison} & $\mathbf{\mediandiff}$ & $\mathbf{D_{KS}}$ & $\mathbf{D_{KL}}$\\
\hline
$\experiment$ vs. $\inter$ & 0.0431 & 0.0085 & 0.0152 \\
$\experiment$ vs. $\nonint$ & 0.0352 & 0.0128 & 0.0135 \\
$\inter$ vs. $\nonint$ & 0.0078 & 0.0057 & 0.0035 \\
\hline
\end{tabular}
\begin{flushleft}By measures of the difference between median values $\mediandiff$, the Kolmogorov-Smirnov distance $D_{KS}$, and the Kullback-Leibler divergence $D_{KL}$, the cumulative distributions for $\inter$, $\nonint$, and $\experiment$ are nearly identical. Since $\ann$ is an angle measured in radians, the values of $\mediandiff$ should be compared to the value $2\pi$. The three distributions are shown in Fig.\,\ref{fig:group}(B).
\end{flushleft}
\label{tab:statsann}
 \end{table}

\begin{table}[!ht]
\caption{
\bf Measures comparing cumulative distributions of fraction of aphids moving $\boldsymbol{\fracmoving}$ in experiment ($\boldsymbol{\experiment}$), a social interaction model ($\boldsymbol{\inter}$) and a noninteracting control model ($\boldsymbol{\nonint}$).}
\begin{tabular}{lccc}
\hline
{\bf Comparison} & $\mathbf{\mediandiff}$ & $\mathbf{D_{KS}}$ & $\mathbf{D_{KL}}$\\
\hline
$\experiment$ vs. $\inter$ & 0.0774 & 0.3226 & 0.7789 \\
$\experiment$ vs. $\nonint$ & 0.4711 & 0.8915 & 0.8373 \\
$\inter$ vs. $\nonint$ & 0.3938 & 0.8806 & 1.6649 \\
\hline
\end{tabular}
\begin{flushleft}By measures of the difference between median values $\mediandiff$, the Kolmogorov-Smirnov distance $D_{KS}$, and the Kullback-Leibler divergence $D_{KL}$, the cumulative distribution of $\inter$ comes closer to $\experiment$ than $\nonint$ does. This is especially apparent in the $\mediandiff$ and $D_{KS}$ values. The three distributions are shown in Fig.\,\ref{fig:group}(C).
\end{flushleft}
\label{tab:statspercent}
 \end{table}

\end{document}